\documentclass[%
reprint,
superscriptaddress,
showpacs,
amsmath,
amssymb,
aps,
]{revtex4-2}
\usepackage{graphicx}
\usepackage{dcolumn}
\usepackage{bm}
\usepackage{CJKutf8}
\usepackage{color}
\usepackage{amssymb}
\usepackage{amsfonts}
\usepackage[colorlinks,urlcolor=blue,linkcolor=blue,citecolor=blue,anchorcolor=blue]{hyperref}
\usepackage{palatino}
\begin{document}
	
	\title{Observation of nonlinear fractal higher-order topological insulator}
	
	\author{Victor O. Kompanets}
	\email[These two authors contributed equally to this work.]{}
	\affiliation{Institute of Spectroscopy, Russian Academy of Sciences, Troitsk, Moscow 108840, Russia}%
	
	\author{Hua Zhong}
	\email[These two authors contributed equally to this work.]{}
	\affiliation{Key Laboratory for Physical Electronics and Devices, Ministry of Education, School of Electronic Science and Engineering, Xi'an Jiaotong University, Xi'an 710049, China}%
	
	\author{Yiqi Zhang}
	\email{zhangyiqi@xjtu.edu.cn}
	\affiliation{Key Laboratory for Physical Electronics and Devices, Ministry of Education, School of Electronic Science and Engineering, Xi'an Jiaotong University, Xi'an 710049, China}%
	
	\author{Yaroslav V. Kartashov}
	\email{kartashov@isan.troitsk.ru}
	\affiliation{Institute of Spectroscopy, Russian Academy of Sciences, Troitsk, Moscow 108840, Russia}%
	
	\author{Yongdong Li}
	\affiliation{Key Laboratory for Physical Electronics and Devices, Ministry of Education, School of Electronic Science and Engineering, Xi'an Jiaotong University, Xi'an 710049, China}%
	
	\author{Sergei A. Zhuravitskii}
	\affiliation{Institute of Spectroscopy, Russian Academy of Sciences, Troitsk, Moscow 108840, Russia}%
	\affiliation{Quantum Technology Centre, Faculty of Physics, M. V. Lomonosov Moscow State University, Moscow 119991, Russia}%
	
	\author{Nikolay N. Skryabin}
	\affiliation{Institute of Spectroscopy, Russian Academy of Sciences, Troitsk, Moscow 108840, Russia}%
	\affiliation{Quantum Technology Centre, Faculty of Physics, M. V. Lomonosov Moscow State University, Moscow 119991, Russia}%
	
	\author{Ivan V. Dyakonov}
	\affiliation{Quantum Technology Centre, Faculty of Physics, M. V. Lomonosov Moscow State University, Moscow 119991, Russia}%
	
	\author{Alexander A. Kalinkin}
	\affiliation{Institute of Spectroscopy, Russian Academy of Sciences, Troitsk, Moscow 108840, Russia}%
	\affiliation{Quantum Technology Centre, Faculty of Physics, M. V. Lomonosov Moscow State University, Moscow 119991, Russia}%
	
	\author{Sergei P. Kulik}
	\affiliation{Quantum Technology Centre, Faculty of Physics, M. V. Lomonosov Moscow State University, Moscow 119991, Russia}%
	
	\author{Sergey V. Chekalin}
	\affiliation{Institute of Spectroscopy, Russian Academy of Sciences, Troitsk, Moscow 108840, Russia}%
	
	\author{Victor N. Zadkov}
	\affiliation{Institute of Spectroscopy, Russian Academy of Sciences, Troitsk, Moscow 108840, Russia}%
	\affiliation{Faculty of Physics, Higher School of Economics, Moscow 105066, Russia}%
	
	
	\begin{abstract}
		\noindent Higher-order topological insulators (HOTIs) are unique materials hosting topologically protected states, whose dimensionality is at least by a factor of 2 lower than that of the bulk. Topological states in such insulators may be strongly confined in their corners that leads to considerable enhancement of nonlinear processes involving such states. However, all nonlinear HOTIs demonstrated so far were built on periodic bulk lattice materials. Here we demonstrate first \textit{nonlinear photonic} HOTI with the fractal origin. Despite their fractional effective dimensionality, the HOTIs constructed here on two different types of the Sierpi\'nski gasket waveguide arrays, may support topological corner states for unexpectedly wide range of coupling strengths, even in parameter regions where conventional HOTIs become trivial. We demonstrate thresholdless solitons bifurcating from corner states in nonlinear fractal HOTIs and show that their localization can be efficiently controlled by the input beam power. We observe sharp differences in nonlinear light localization on outer and multiple inner corners and edges representative for these fractal materials. Our findings not only represent a new paradigm for nonlinear topological insulators, but also open new avenues for potential applications of fractal materials to control the light flow.
	\end{abstract}
	
	\maketitle
	
	Fractals are self-similar structures (i.e., next generation of a fractal can be constructed by combining copies of its previous generation) widely represented in universe~\cite{bunde.book.1994}, whose unusual internal composition finds its manifestation in new physical phenomena observed in solid-state physics, acoustics, and photonics, to mention just a few areas~\cite{song.apl.104.233106.2014, he.epl.112.17010.2015, kempkes.np.15.127.2019, pai.prb.100.155135.2019, iliasov.prb.101.045413.2020, fremling.prr.2.013044.2020, xu.np.15.703.2021, liu.prl.126.176102.2021, ivaki.cp.5.327.2022, manna.pra.105.L021302.2022, yang.light.9.128.2020, biesenthal.science.376.1114.2022, zheng.sb.67.2069.2022, li.sb.67.2040.2022, ren.nano.12.3829.2023}. 
	One of the distinguishing characteristics of a fractal is its fractional dimension, which can be described by the non-integer effective Hausdorff dimension ${d_f = \log_\ell m }$, where $m$ is the number of previous-generation elements required to construct next-generation fractal, while $\ell$ is the factor, by which length of the fractal edge would increase in the next generation. Thus, famous Sierpi\'nski carpet and gasket fractals have Hausdorff dimensions ${d_f = \log_3 8}$ and ${d_f = \log_2 }3$, respectively, which reflects their different composition. Fractal systems are aperiodic, but regular---they are sometimes considered as lacking ``bulk" due to the presence of multiple holes, inner edges and corners. In photonics, such unusual composition of fractal structures may open new prospects for manipulation and localization of the light fields in them, including realization of quantum anomalous transport~\cite{xu.np.15.703.2021} and flat-band systems~\cite{xie.aplp.6.116104.2021, song.lpr.17.2200315.2023}. Particularly intriguing problem is the possibility of realization of topological phases in fractal structures, since the appearance of topological edge states is tightly connected with the dimensionality of the system and usually stems from topological properties of its bulk, which can be very specific in fractal systems.

	Photonic systems offer a unique testbed for realization of topologically nontrivial structures~\cite{lu.np.8.821.2014, ozawa.rmp.91.015006.2019, zhang.nature.618.687.2023}, including various types of Chern~\cite{wang.nature.461.772.2009}, Floquet~\cite{rechtsman.nature.496.196.2013}, valley-Hall~\cite{noh.prl.120.063902.2018},
	and higher-order~\cite{peterson.nature.555.346.2018, noh.np.12.408.2018, mittal.np.13.692.2019, hassan.np.13.697.2019, xie.nrp.3.520.2021,lin.nrp.5.483.2023}
	topological insulators. Most of these systems, including HOTIs, were constructed on structures with periodic bulk. However, very recently it was shown that unidirectional travelling topological edge states can form in fractal waveguide arrays with helical channels~\cite{yang.light.9.128.2020, biesenthal.science.376.1114.2022} and in fractal Haldane model~\cite{li.prr.5.023189.2023} implying that formal absence of insulating bulk in fractals is not an obstacle for realization of topological phase, and bulk-edge correspondence~\cite{hatsugai.prl.71.3697.1993} is still meaningful for these systems. While fractal HOTIs were proposed in electronic systems~\cite{manna.prb.105.L201301.2022, brzezinska.prb.98.205116.2018} and recently realized in acoustics~\cite{zheng.sb.67.2069.2022, li.sb.67.2040.2022}, higher-order topological states in \textit{photonic fractals} were never observed so far. At the same time, first experiments in acoustics hint on a very unusual manifestation of topological effects in fractal systems~\cite{zheng.sb.67.2069.2022, li.sb.67.2040.2022}, connected with the possibility of localization in their multiple inner corners, and strong dependence of the parameter range, where fractal system is topological on fractal generation order, making them clearly distinct from conventional HOTIs.

	Among the advantages of photonic systems in comparison with the electronic and acoustic ones, is that the former systems can be strongly nonlinear. Nonlinearity not only offers a convenient knob for controlling localization and propagation dynamics of the topological excitations, but it is often crucial for the effects that determine practical applications of such systems \cite{smirnova.apr.7.021306.2020}, ranging from lasing and harmonic generation to bistability,  nonlinearity-controlled switching and routing with th topologically protected states. In addition, nonlinearity gives rise to a broad spectrum of topological edge solitons, inheriting topological protection from their linear counterparts~\cite{lumer.prl.111.243905.2013, ablowitz.pra.90.023813.2014, leykam.prl.117.143901.2016, kartashov.optica.3.1228.2016, ablowitz.pra.96.043868.2017, ivanov.acs.7.735.2020, zhong.ap.3.056001.2021} as demonstrated in~\cite{mukherjee.science.368.856.2020, mukherjee.prx.11.041057.2021, xia.light.9.147.2020, guo.ol.45.6466.2020, kartashov.prl.128.093901.2022}, and it may create the self-induced topological phases \cite{hadad.ne.1.178.2018, zangeneh.prl.123.053902.2019, maczewsky.science.370.701.2020}. Nonlinear effects and formation of the unique corner solitons in HOTIs were recently reported too~\cite{kirsch.np.17.995.2021, hu.light.10.164.2021, arkhipova.sb.68.2017.2023}. Nowadays, there is a considerable interest to investigation of the nonlinear effects in aperiodic topological photonic systems, with only a few observations available so far~\cite{ren.light.12.194.2023}. Fractal topological photonic systems may thus provide a unique platform for investigation of such effects.

	In this Article, we describe first experimental realization of photonic fractal HOTI and study the interplay between topological and nonlinear effects in this aperiodic system, which gives rise to topological corner solitons. To demonstrate such states we utilize fractal Sierpi\'nski gasket waveguide arrays of two different types inscribed in fused silica using fs laser writing technique. Higher-order topological phase is realized due to the controllable shift of the waveguides that adjusts coupling strengths between sites in the first and subsequent generations of fractals and is manifested in the appearance of the corner states of the topological origin. The remarkable distinctive property of this photonic system is that topological states can appear not only in the outer, but also in the inner corners of the structure. Moreover, topological corner states here can exist even for the shifts, where conventional HOTIs with periodic bulk become trivial. To characterize topological properties of this aperiodic system we employ real-space polarization index~\cite{wheeler.prb.100.245135.2019,kang.prb.100.245134.2019}. Finally, we demonstrate thresholdless fractal topological corner solitons bifurcating from their linear counterparts and existing in the forbidden spectral gaps. 
	
	\begin{figure*}[htbp]
		\centering
		\includegraphics[width=1\textwidth]{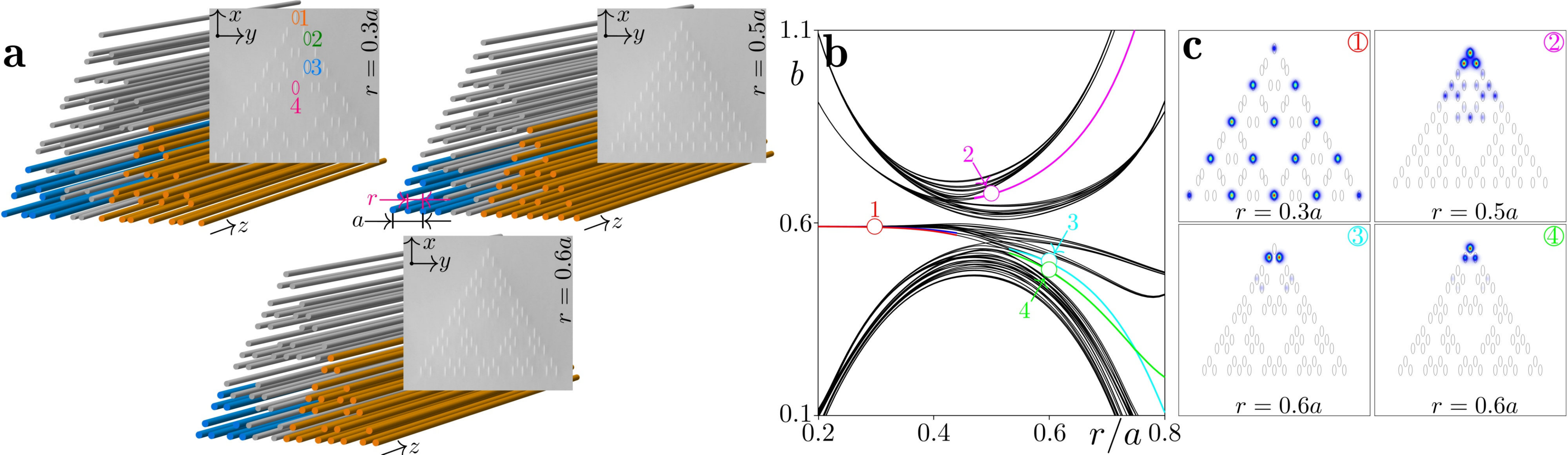}
		\caption{Fractal higher-order topological insulators and their linear spectra.
			(\textbf{a}) Schematic representation of the third generation $G_3$ of the Sierpi\'nski gasket waveguide arrays with ${r=0.3a}$, $0.5a$, and $0.6a$. Blue and orange sectors of these arrays represent the first-generation $G_1$ and second-generation $G_2$ structures, respectively. Microphotographs of the corresponding fs-laser written waveguide arrays are presented in the insets. The orange, green, blue, and magenta ellipses in microphotographs indicate the representative sites 1, 2, 3, and 4 that will be used below for probing of excitation dynamics. (\textbf{b}) Eigenvalues $b$ of the stationary linear states of the fractal array vs distortion parameter $r$. Colored curves represent localized states, while black ones correspond to the delocalized states. (\textbf{c}) Intensity distributions shown within ${-23\le x,y\le 23}$ window for four representative eigenstates corresponding to the colored circles in (\textbf{b}). Here and in figures below the array depth is ${p=5.7}$.}
		\label{fig1}
	\end{figure*}

	\section*{Results and Discussions}
	\subsection*{Fractal arrays, their linear spectra and eigenmodes}
	
	We consider two different types of Sierpi\'nski gasket arrays in this Article, which are further termed case-1 and case-2 arrays. Both arrays are produced using identical first-generation triangular element $G_1$ highlighted with the blue color in the schematic representation in Fig.~\ref{fig1}\textbf{a}, where we show third generation $G_3$ of case-2 structure (for more details of fractal array construction and results on case-1 structure see \textbf{Supporting Information}). $G_n$ generation of the Sierpi\'nski gasket fractal array is formed by three $G_{n-1}$ generation structures (for instance, $G_2$ elements in Fig.~\ref{fig1}\textbf{a} are highlighted with orange color)---on this reason fractal arrays are self-similar. In case-2 arrays these previous generations share three common sites, while in case-1 array there are no such common sites. Consequently, $G_n$ generation includes $3^{n+1}$ sites in case-1 array, and ${3^{n+1}-(3^n-3)/2}$ sites in case-2 array. Due to the method of their construction, fractal arrays posses multiple holes, inner corners and edges. They are characterized by the effective Hausdorff dimension ${d_f = \log_2 }3$ that is lower than $2$. Further we focus on newly designed case-2 arrays, discussing solitons in the case-1 arrays in \textbf{Supporting Information}. To realize fractal HOTI, we introduce controllable distortion (via the parameter $r$) into the structure by shifting the neighboring waveguides in the opposite directions, while keeping spacing $a$ between the next-nearest-neighbor waveguides constant, as indicated in Fig.~\ref{fig1}\textbf{a}. The examples of undistorted (${r=0.5a}$) and distorted (${r=0.3a}$ and ${r=0.6a}$) arrays are shown in Fig.~\ref{fig1}\textbf{a} with the photographs of such fs-laser written entities in fused silica.
	
	Propagation of light beams in fractal arrays inscribed in focusing cubic medium is described by the nonlinear Schr\"odinger equation:
	\begin{equation}\label{eq1}
		i \frac{\partial \psi}{\partial z}=-\frac{1}{2} \left( \frac{\partial^2}{\partial x^2} + \frac{\partial^2}{\partial y^2} \right) \psi
		-\mathcal{R}(x,y) \psi-|\psi|^{2} \psi,
	\end{equation}
	where $\psi$ is the dimensionless complex amplitude of the light field; $x,y$ and $z$ are the normalized transverse coordinates and propagation distance, respectively; the function $ {\mathcal{R}(x,y) = p \sum_{mn} e^{-(x-x_m)^2/d_x^2 - (y-y_n)^2/d_y^2}}$ describes the fractal case-2 waveguide array composed of single-mode elliptical (due to writing process) waveguides with the depth $p$ and widths $d_{x,y}$ placed in the nodes $(x_m,y_n)$ of the Sierpi\'nski gasket grid. Further, we use arrays with ${p=5.7}$, ${a=6.0}$, and ${d_x=0.25}$, ${d_y=0.75}$ corresponding to the parameters of the experimental structures (see \textbf{Materials and Methods} for details of normalization).
	
	We first characterize linear spectrum of such arrays by omitting nonlinearity in Eq.~(\ref{eq1}) and calculating the linear eigenmodes of the form ${\psi=u(x,y)e^{ibz}}$, where $u$ describes modal shape, and $b$ is the propagation constant (eigenvalue). Linear spectrum of fractal array of third generation $G_3$ in the form of dependence of eigenvalues of all supported modes $b$ on distortion parameter $r$ is presented in Fig.~\ref{fig1}\textbf{b}, where colored curves correspond to the localized states, while black curves correspond to the extended states. Remarkably, in fractal arrays localized in-gap states are encountered in both ${r>0.5a}$ and ${r<0.5a}$ regimes, in sharp contrast with HOTIs with periodic bulk, where localized states appear only for one of these types of distortion leading to dominance of the inter-cell coupling over the intra-cell one. Notice that the state shown with magenta curve can exist even at ${r=0.5a}$. Intensity distributions of four typical localized states corresponding to circles in Fig.~\ref{fig1}\textbf{b} are shown in Fig.~\ref{fig1}\textbf{c}. Magenta, green, and cyan branches correspond to the co-existing \textit{outer corner states} with different internal structure, representative namely for the case-2 fractal array. In contrast, red branch corresponds to the states, where strongly localized spots appear simultaneously in multiple inner and outer corners, so one can call them \textit{hybrid corner states}, as they reflect unique internal composition of the fractal array (there are several such branches in spectrum with spots only in the corners that become nearly degenerate for small $r$ values). Outer corner states are all three-fold degenerate, while their number is not affected by the fractal generation order $n$. According to the definition of the effective dimensionality ${d_e=\lim_{n\to\infty}{\ln N/\ln N_l}}$, where $N$ is the number of sites occupied by the corner state and ${N_l=3\cdot 2^{n-1}+1}$ is the total number of sites on one outer edge~\cite{zheng.sb.67.2069.2022, li.sb.67.2040.2022}, outer corner states are zero-dimensional, as ${d_e \to 0}$ with increase of the fractal generation order $n$. In contrast, the number of spots in hybrid corner states ${N=(3^n+3)/2}$ increases with $n$, so that $d_e$ for hybrid state approaches effective Hausdorff dimensionality $d_f=\log_23$ of the structure. Thus, fractal HOTIs offer unique opportunity to observe photonic corner states of the different effective dimensionality.
	
	To characterize topological properties of fractal Sierpi\'nski gasket waveguide arrays one resorts to the real-space polarization index~\cite{wheeler.prb.100.245135.2019, kang.prb.100.245134.2019}. The remarkable prediction of this analysis (we glue two Sierpi\'nski gaskets to form a rhombic structure---see \textbf{Materials and Methods} and \textbf{Supporting Information} for details) is that the index for both hybrid and outer corner states are $0.5$ in their respective existence domains, meaning that all such corner states are of topological origin. Similar conclusions can be drawn for the case-1 fractal arrays, as we also discuss in \textbf{Supporting Information}. Thus, fractal Sierpi\'nski gasket arrays allow to realize HOTIs in a broader range of distortion parameters $r$ in comparison with HOTIs based on the periodic kagome and Su-Schrieffer-Heeger arrays~\cite{kirsch.np.17.995.2021, hu.light.10.164.2021}, where in our notations corner states are possible only at ${r<0.5a}$.
	
	\subsection*{Topological soliton families}
	
	\begin{figure*}[htbp]
		\centering
		\includegraphics[width=1\textwidth]{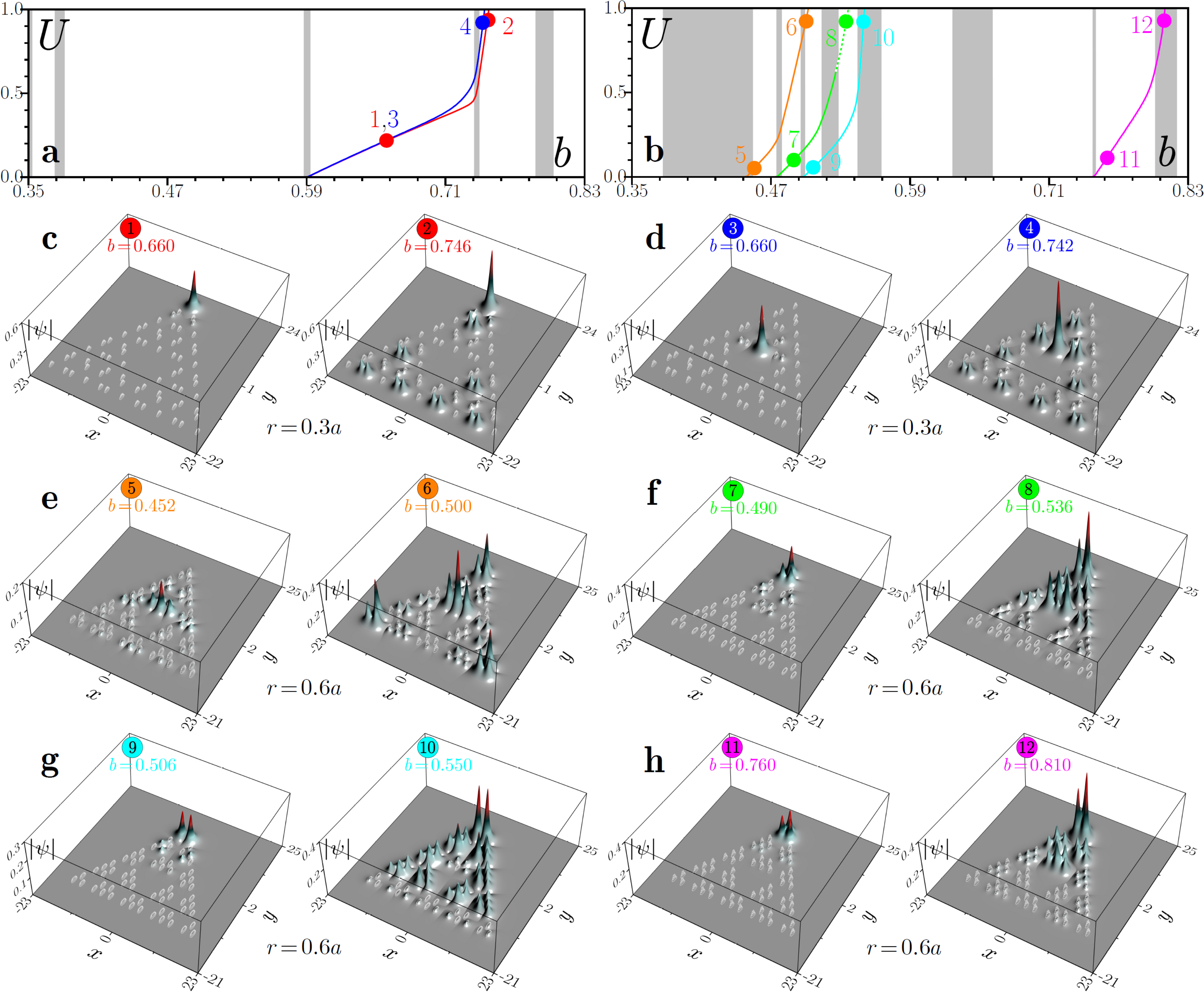}
		\caption{Topological solitons in fractal insulators.
			(\textbf{a}) Power vs propagation constant for the solitons bifurcating from the linear eigenstates concentrated around sites 1 (red curve) and 4 (blue curve) in the fractal array with ${r=0.3a}$. Panels (\textbf{c}) and (\textbf{d}) show field modulus distributions corresponding to the dots in (\textbf{a}). (\textbf{b}) Power vs propagation constant for the solitons bifurcating from linear eigenstates concentrated around sites 1 (green, cyan and magenta curves) and 4 (orange curve) in fractal array with ${r=0.6a}$. Field modulus distributions corresponding to the dots in (\textbf{b}) are shown in (\textbf{e})--(\textbf{h}). In (\textbf{a}) and (\textbf{b}) gray regions are associated either with the eigenvalues of the localized states or with the bands of the delocalized states.}
		\label{fig2}
	\end{figure*}
	
	In the presence of nonlinearity localized corner states in fractal arrays give rise to the families of bifurcating from them thresholdless topological corner solitons. Such solitons also have the form ${\psi=u(x,y)e^{ibz}}$, whose substitution into Eq. (\ref{eq1}) yields
	${bu = \frac{1}{2} ( \partial^2_x + \partial^2_y ) u + \mathcal{R} u + u^3}$
	that was solved using Newton method. Now propagation constant $b$ parameterises the family of solitons, determining their power ${U = \int |\psi|^2 dx dy}$.
	
	The families of corner solitons in fractal insulators can be very rich. In Fig.~\ref{fig2} we show representative $U(b)$ dependencies in $G_3$ structures. At ${r=0.3a}$, when a set of practically degenerate linear topological modes with spots localized only on the inner and outer corners exist, topological solitons can form in any corner of fractal array, see red (outer corner) and blue (inner corner) families in Fig.~\ref{fig2}\textbf{a} bifurcating from corresponding linear modes (or their linear combination, because for this $r$ there exist a set of practically degenerate \textit{hybrid corner states}). In this figure wider gray regions correspond to the bands occupied by extended states in linear spectrum, while narrow gray lines correspond to the eigenvalues of linear in-gap corner states. For the selected $r$ value such solitons are strongly localized at the low and intermediate powers (see profiles in Fig. \ref{fig2}\textbf{c} and \ref{fig2}\textbf{d} corresponding to points $1$,$3$ in Figs.~\ref{fig2}\textbf{a}). For illustrative purposes here we superimpose $|\psi|$ distribution for soliton on array profile $\mathcal{R}$ shown by white ellipses. Importantly, because nonlinearity leads to shift of the propagation constant within the gap of linear spectrum, one can control localization degree of corner solitons by increasing their power. Namely, such solitons gradually extend, developing long tails as their propagation constant approaches nearest band of extended states, eventually coupling with them and loosing localization, when $b$ enters into such band (as illustrated by profiles in the points $2,4$ of Fig.~\ref{fig2}\textbf{a}, where soliton clearly couples with the edge states).
	
	\begin{figure*}[htbp]
		\centering
		\includegraphics[width=1\textwidth]{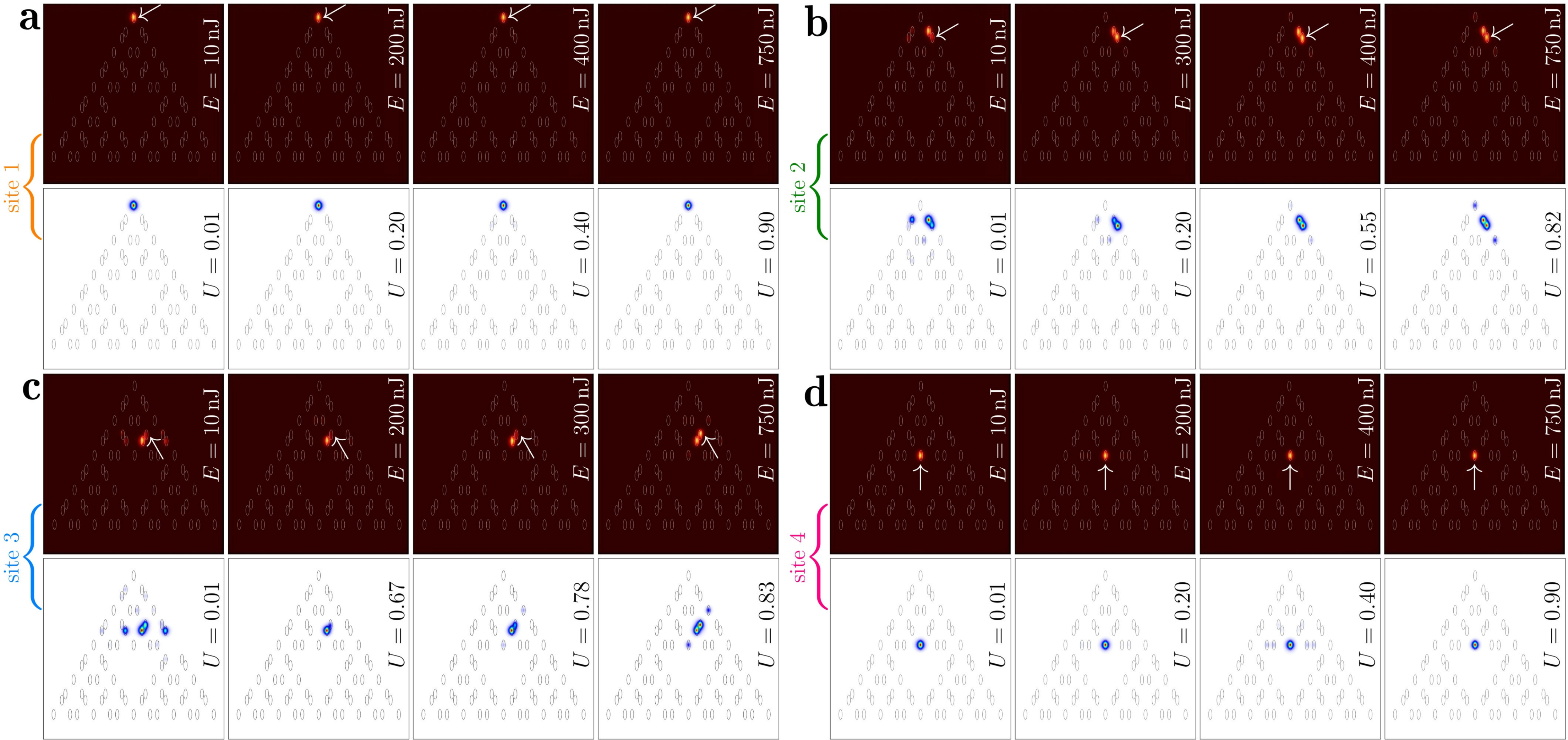}
		\caption{Observation of nonlinear states in fractal array with ${r=0.3a}$. Output intensity distributions after 10 cm of propagation for the excitation of sites 1 (\textbf{a}), 2 (\textbf{b}), 3 (\textbf{c}) and 4 (\textbf{d}) in $G_3$ fractal array. Figures with the maroon background show experimental results, while figures with the white background show results of theoretical simulations. Gray ellipses in all panels indicate waveguide locations. Input pulse energies $E$ and powers $U$ are indicated on each plot. All distributions are shown within the window ${-23\le x,y\le 23}$. Arrows in the experimental panels indicate position of the input excitation.}
		\label{fig3}
	\end{figure*}
	
	Fractal HOTI with ${r=0.6a}$ supports multiple families of topological solitons with different symmetries forming in \textit{outer corners}. They are shown in Fig.~\ref{fig2}\textbf{b} with magenta, cyan, and green lines, in accordance with colors of linear corner states in Fig.~\ref{fig1}\textbf{b}, from which these solitons bifurcate. Solitons inherit symmetry of the corresponding linear states. Thus, in green family (Fig.~\ref{fig2}\textbf{f}, points $7$,$8$) corner maximum is out-of-phase with maxima in two nearest-neighbour sites; solitons from blue family have empty outer corner site, while fields in nearest-neighbor sites are out-of-phase (Fig.~\ref{fig2}\textbf{g}, points $9$,$10$); while most localized solitons from magenta branch have nearly equal intensities in three close sites in the corner (Fig.~\ref{fig2}\textbf{h}, points $11$,$12$). These solitons typically gradually broaden with increase of power and strongly expand into array when coupling with extended states occurs upon entering of their propagation constant $b$ into different bands of the linear spectrum. Notice that extended states occupying multiple sites can also produce nonlinear families, as illustrated by orange branch in Fig.~\ref{fig2}\textbf{b}, see profiles in Fig.~\ref{fig2}\textbf{e} (the power of this state vanishes in the bifurcation point only because our array is finite). Interestingly, with increase of the power this extended state couples with the outer corner states (point \textbf{6}, Fig.~\ref{fig2}\textbf{e}).
	
	Despite large variety of corner solitons appearing in fractal HOTI, nearly all of them are dynamically stable. Their stability was verified by adding small-scale random perturbation ($5\%$ in the amplitude) into input field distributions and propagating them in the frames of Eq.~(\ref{eq1}) over a very long distances ${z\sim 10^4}$. For all branches shown in Figs.~\ref{fig2}\textbf{a} and \ref{fig2}\textbf{b} with solid lines, such perturbations resulted only in small-amplitude oscillations signalizing on their stability, while decay was observed only for a small part of the green branch shown with the dashed line.
	
	\subsection*{Observation of the corner solitons in fractal HOTI}
	
	To demonstrate corner solitons in fractal HOTIs we fabricated a set of case-2 $G_3$ structures with various distortion parameters ${r = 0.3a}$, ${r=0.5a}$ and ${r=0.6a}$ using fs-laser inscription technique (for experiments with case-1 array see \textbf{Supporting Information}). Arrays were inscribed in $10~\textrm{cm}$-long fused silica samples (see \textbf{Materials and Methods} for details of fabrication). Exemplary photographs of the inscribed arrays are presented as insets in Fig.~\ref{fig1}\textbf{a}. We selected four representative locations to study excitation dynamics, as indicated with colored ellipses with numbers 1 (outer corner site), 2 (site on the outer edge), 3 (site on the inner edge site) and 4 (inner corner site). For excitation we used pulses with the duration about $280\,\textrm{fs}$ of variable energy $E$ derived from $1\,\textrm{kHz}$ Ti:sapphire laser to achieve strong nonlinear response (see \textbf{Materials and Methods}).	
	
	\begin{figure*}[htbp]
		\centering
		\includegraphics[width=1\textwidth]{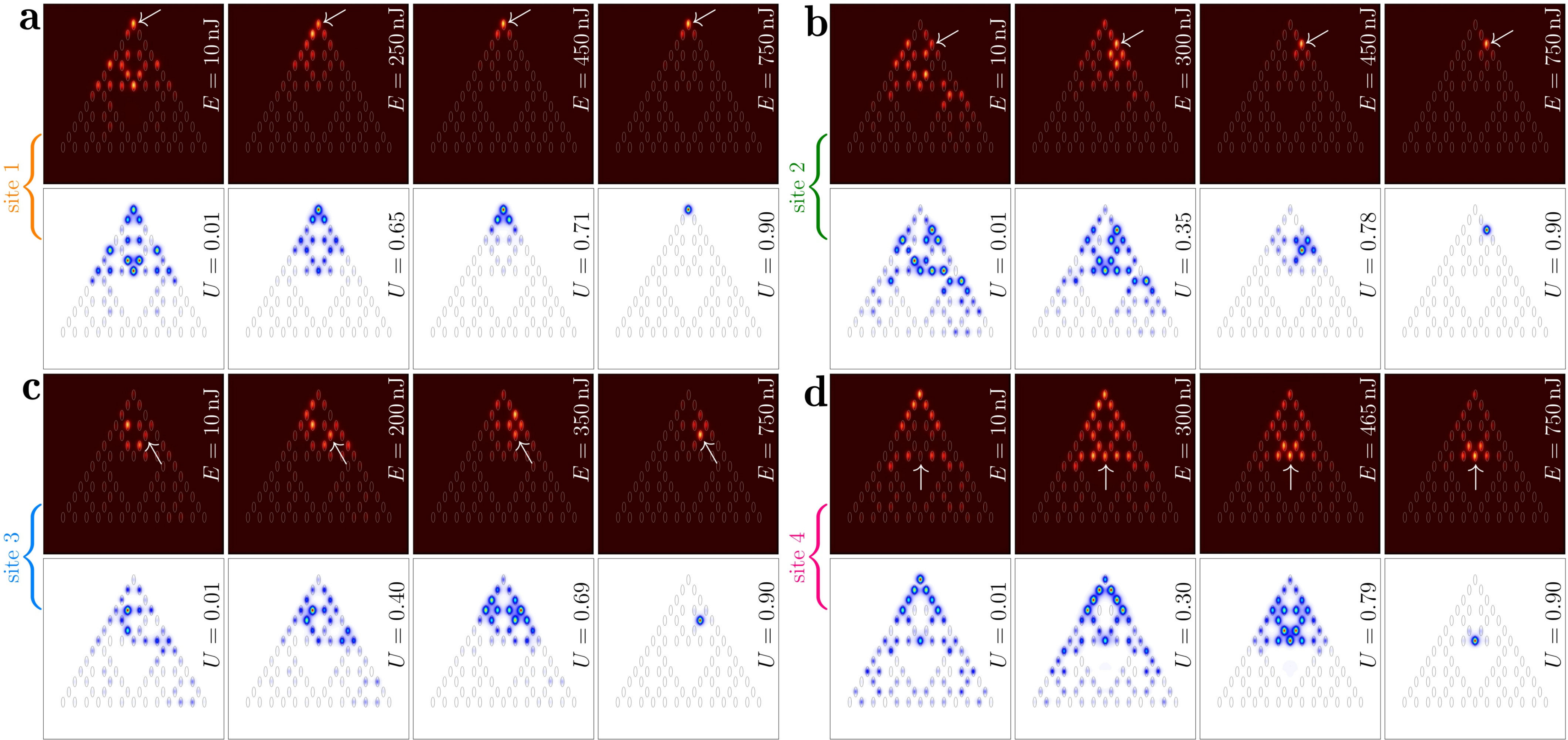}
		\caption{Nonlinear dynamics in fractal array with ${r=0.5a}$. Rich nonlinear dynamics and gradual transition to the localization at the highest power levels as observed for the excitations of sites 1-4 in $G_3$ fractal array. The arrangement of panels is the same as in Fig.~\ref{fig3}. Arrows in experimental panels indicate position of the input excitation.}
		\label{fig4}
	\end{figure*}
	
	First, we investigate the structure with ${r=0.3a}$ that supports \textit{hybrid corner states}. In Figs.~\ref{fig3}\textbf{a}-\ref{fig3}\textbf{d}, we compare experimental output intensity distributions (images with maroon background) obtained at the different pulse energies $E$ for the excitation of four above mentioned sites of the array (in each case, the excited sites are indicated by white arrows) with theoretical distributions for the different powers $U$ obtained by solving Eq.~(\ref{eq1}) (images with white background). Since well-localized thresholdless solitons at $r=0.3a$ can form in any inner or outer corner of this structure, the excitation of sites 1 and 4 (Figs.~\ref{fig3}\textbf{a} and \ref{fig3}\textbf{d}) yields strongly localized, practically single-site patterns in both linear (${E=10\,\rm nJ}$) and nonlinear (e.g., ${E=750\,\rm nJ}$) regimes, confirming the formation of thresholdless corner solitons. In contrast, when exciting edge sites 2 and 3 (Figs.~\ref{fig3}\textbf{b} and \ref{fig3}\textbf{c}), we observe diffraction and dynamic oscillations of power between close pairs of waveguides even at pulse energies ${E \sim 800\,\rm nJ}$, indicating that inner and outer edges do not support well-localized thresholdless in-gap states and significant power levels are needed to achieve localization at such edges.
	
	Turning to array with ${r=0.5a}$, where spacing between nearest sites is the same in the entire array, one should take into account that even though magenta branch of outer corner states already exists in this borderline case, its localization is rather weak (see state 2 in Fig.~\ref{fig1}\textbf{c}), thus efficiency of its excitation with single-site input is low. On this reason, in Fig.~\ref{fig4}\textbf{a} showing experimental patterns for this structure, the beam with ${E=10\,\rm nJ}$ launched into site 1 experiences diffraction, even though some fraction of power clearly remains in the corner waveguide. Diffraction is nearly suppressed when the pulse energy increases to the moderate value of ${E=250\,\rm nJ}$, while further increase of $E$ results in the excitation of the outer corner soliton. For excitation of sites 2, 3, and 4 shown in Figs.~\ref{fig4}\textbf{b}, \ref{fig4}\textbf{c} and \ref{fig4}\textbf{d}, respectively, one needs substantially higher pulse energies to achieve comparable degree of localization, while particularly for sites 3 and 4 localization remains weak even for the energies of ${E \sim 750\,\rm nJ}$.
	
	\begin{figure*}[htbp]
		\centering
		\includegraphics[width=1\textwidth]{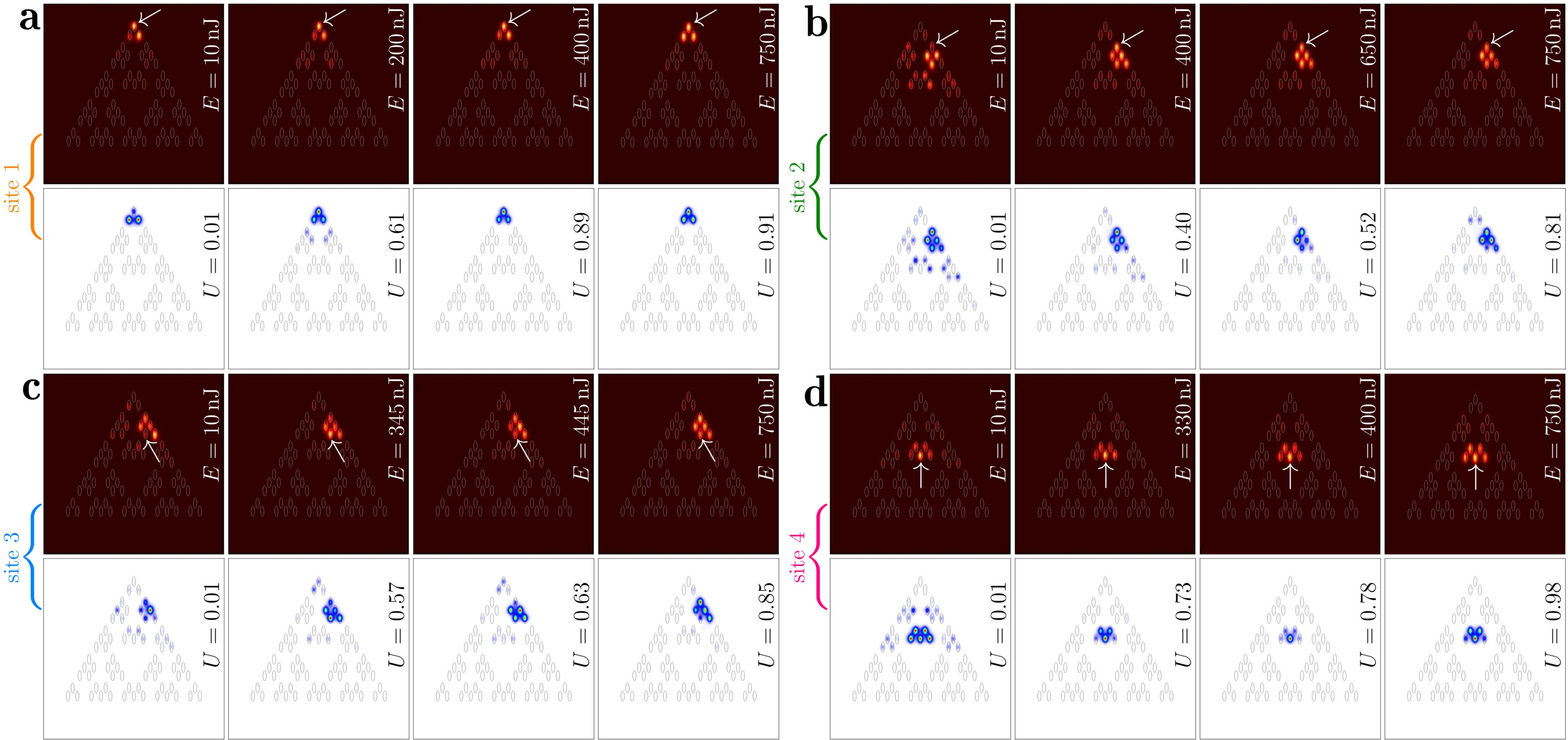}
		\caption{Observation of nonlinear states in fractal array with ${r=0.6a}$. Formation of the localized thresholdless nonlinear states for the excitation of sites 1 (\textbf{a}) and 4 (\textbf{d}) and rich nonlinear dynamics with the light switching between several closely located waveguides for excitation of sites 2 (\textbf{b}) and 3 (\textbf{c}) observed in $G_3$ fractal array. Arrows in the experimental panels indicate position of the input excitation.}
		\label{fig5}
	\end{figure*}
	
	Fractal case-2 array with ${r=0.6a}$ supports three types of outer corner solitons with different symmetries, as shown in Fig.~\ref{fig2}\textbf{b}. Among them, the soliton belonging to cyan branch cannot be excited by focusing beam into site 1, because this site is empty for such nonlinear states (due to their parity). Therefore, excitation of site 1 is supposed to yield nonlinear combination of states from magenta and green branches of Fig.~\ref{fig2}\textbf{b}. This is what we observe in Fig.~\ref{fig5}\textbf{a}, where for all pulse energies from $10$ to ${750\,\rm nJ}$ light remains practically confined in three closely spaced the outer corner waveguides, closely resembling profile of the outer corner soliton and confirming that in this case it is also thresholdless. For ${r=0.6a}$ no localized linear states exist in the array, except for states in outer corners. Consequently, excitations of sites 2 and 3 at all pulse energies yield broad output distributions, where power oscillates between five waveguides, and where intensity maximum may not be located in the excited waveguide (Figs.~\ref{fig5}\textbf{b} and \ref{fig5}\textbf{c}). Similarly, excitation of site 4 yields nonlinear state akin to nontopological delocalized state from orange branch of Fig.~\ref{fig2}\textbf{b}, whose expansion with increase of $E$ is obvious from experimental patterns in Fig.~\ref{fig5}\textbf{d} (compare this state with well-localized inner corner soliton obtained in the same location at ${r=0.3a}$).
	
	\section*{Conclusion}
	
	In summary, we have reported the first example of nonlinear photonic fractal HOTI that supports a rich variety of topological corner states. The remarkable new feature of fractal structures considered here is that they possess corner states (that may have different effective dimensionality) for a very broad range of distortion parameters, substantially exceeding the range, where higher-order topological phase emerges in HOTIs built on periodic lattices. The presence of such states facilitates nonlinear light localization and resulted in observation of thresholdless corner solitons, in both outer and inner corners of these structures. Our results extend the class of HOTIs and highlight new prospects for exploration and practical utilization of nonlinear phenomena in photonic fractals. They may be used in new designs of topological lasers that can potentially emit in richer set of states than conventional higher-order topological lasers (for example, depending on gain landscape and amplitude, lasing can occur either in outer or in inner corners of the structure, or in corner states with different parity), in shaping of higher-harmonic fields in various parametric processes, design of fractal microresonator networks and quantum interfaces of fractal structures, control of condensation in light-matter systems with strong coupling, like polariton condensates in fractal microcavities under resonant or nonresonant pump, and in many other settings.

	\section*{Materials and Methods}
	
	\subsection*{Normalization of parameters in theoretical model}
	
	The transverse coordinates $x,y$ in Eq.~(\ref{eq1}) are normalized to the characteristic scale ${r_0=10\,\mu\textrm{m}}$, the propagation distance $z$ is normalized to the diffraction length ${kr_0^2\approx1.14\,\textrm{mm}}$ (corresponding to ${z=1}$), where ${k=2\pi n/\lambda}$ is the wavenumber in the medium with unperturbed refractive index $n$ (for fused silica ${n\approx 1.45}$ and the nonlinear refractive index ${n_2\approx 2.7\times 10^{-20}\, \rm m^2/W}$), and ${\lambda=800~\textrm{nm}}$ is the working wavelength. The array depth ${p=k^2r_0^2\delta n/n}$ is proportional to the refractive index contrast $\delta n$ of the waveguides. Thus, in our arrays next-nearest-waveguide distance ${a=6.0}$ corresponds to ${60~\mu\textrm{m}}$, waveguide widths ${d_x=0.25}$, ${d_y=0.75}$ correspond to ${2.5~\mu\textrm{m} \times 7.5~\mu\textrm{m}}$ wide elliptical waveguides, sample length of ${10~\textrm{cm}}$ corresponds to ${z\approx 88}$, while array depth ${p=5.7}$ corresponds to refractive index contrast ${\delta n \approx 6.4\times 10^{-4}}$.
	
	\subsection*{Fs-laser inscription of the waveguide arrays}
	
	Fractal waveguide arrays were written in ${10\,\textrm{cm}}$ long fused silica glass substrate (JGS1). The individual waveguides were inscribed by circularly polarized beam with central wavelength of $515\,\textrm{nm}$, with pulse duration of ${230\,\textrm{fs}}$, repetition rate ${1\,\textrm{MHz}}$, and pulse energy ${270\,\textrm{nJ}}$, focused with an aspheric lens (${\textrm{NA} = 0.3}$) under the sample surface in the depth range from 600 to $1000\,\mu\textrm{m}$ near the preselected optimal depth of ${800\,\mu\textrm{m}}$. Translation of the sample with respect to the focus was performed by a high-precision positioner (Aerotech) with a scanning velocity of ${1\,\textrm{mm/s}}$. Waveguides demonstrate propagation losses less than ${0.3\,\textrm{dB/cm}}$ at ${\lambda = 800\,\textrm{nm}}$.
	
	\subsection*{Experimental excitation of the waveguide arrays}
	
	In experiments, we employed single-waveguide excitations using fs pulses of variable energy $E$ from ${1\,\rm kHz}$ Ti:sapphire laser at ${800\,\rm nm}$ central wavelength. Initially, short pulses with a ${40\,\rm fs}$ duration and wide spectrum from a regenerative amplifier system Spitfire HP (Spectra Physics) first pass through an active beam position stabilization system (Avesta) and an attenuator, and afterwards are launched into a $4f$ single-grating stretcher-compressor with a variable slit. Spectra of such pulses are narrowed by a slit down to ${5\,\rm nm}$, which corresponds to the pulse duration of ${280\,\rm fs}$. This increase in the pulse duration allows to prevent optical collapse and strong spectral broadening during pulse propagation in the waveguides, i.e. it allows to neglect the temporal effects. The pulses after stretcher compressor were focused into selected waveguides and the output intensity distributions after propagation in 10 cm sample were recorded using a Kiralux CMOS camera (Thorlabs). The input peak power in the waveguide (for each pulse in the $1\,\rm kHz$ sequence) was defined as a ratio of the input pulse energy $E$ to the pulse duration ${\tau = 280\,\rm fs}$. Taking into account the losses for the matching with the focusing lens the input power can be evaluated as ${2.5\,\rm kW}$ for each ${1\,\rm nJ}$. For example, maximal excitation energy of ${E = 800\,\rm nJ}$ in experimental patterns presented here corresponds to the peak power of ${2.0\,\rm MW}$. 
	
	\subsection*{Real-space polarization index}
	
	Fractal waveguide arrays are aperiodic structures that are sometimes considered as structures without bulk due to the method of their construction. In such structures the appearance of corner states can be associated with the nonzero real-space polarization, allowing to characterize topological properties of this system:
	${P=-\frac{i}{2 \pi} \ln [\rm{det}(S)]}$,
	where ${S_{m, n}=U_m^{\dagger} e^{i 2 \pi \hat{q} / L} U_n}$, $L$ is the length of the fractal array along the $q$ direction, $\hat q$ is the position operator, $U_n$ is the eigenfunction of $n^\textrm{th}$ state of the fractal array obtained with the periodic boundary conditions in the $q$ direction (that is usually selected along the outer edge of the array). Real-space polarization for the states in our system can be calculated using tight-binding approximation. To be able to apply periodic boundary conditions upon calculation of eigenfunctions $U_n$ we considered therhombic structure composed from two stacked Sierpi\'nski gasket arrays, real and virtual ones, while neglecting coupling with the virtual and missing sites. It was found that real-space polarization index is equal to $0.5$ for both hybrid and outer corner states in their domains of existence in $r$. The detailed information on this method can be found in the \textbf{Supporting Information}.
	
	\section*{Data availability}
	\noindent The data that support the plots within this paper and other findings of this study are available from the corresponding author on a reasonable request.
	
	\section*{Code availability}
	\noindent The analysis codes will be made available on a reasonable request.
	
	\section*{Acknowledgements}
	\noindent This work was supported by the research project FFUU-2021-0003 of the Institute of Spectroscopy of the Russian Academy of Sciences and partially by the Russian Science Foundation (21-12-00096). Yiqi Zhang and Hua Zhong acknowledge funding by the National Natural Science Foundation of China (12074308, 12304370). Sergei A. Zhuravitskii acknowledges support by the Foundation for the Advancement of Theoretical Physics and Mathematics ``BASIS'' (22-2-2-26-1).
	
	\section*{Author contributions}
	\noindent Yiqi Zhang and Yaroslav V. Kartashov formulated the problem.
	Sergei A. Zhuravitskii, Nikolay N. Skryabin, Ivan V. Dyakonov, and Alexander A. Kalinkin fabricated the samples. Victor O. Kompanets, and Sergey V. Chekalin performed experiments. Hua Zhong, Yiqi Zhang and Yongdong Li performed numerical modeling. Yaroslav V. Kartashov, Sergei P. Kulik, and Victor N. Zadkov supervised the work. All co-authors took part in discussion of the results and writing the manuscript.
	
	\section*{Competing interests}
	\noindent The authors declare no competing interests.
	
	\section*{Additional information}
	
	\noindent \textbf{Correspondence and requests for materials} should be addressed to Yiqi Zhang and Yaroslav V. Kartashov.
	
	
	\noindent \textbf{Reprints and permissions} information is available at 
	
	\bibliography{my_library}
	
\end{document}